\documentclass[3p,times, twocolumn]{elsarticle}
\usepackage{color}
\usepackage[normalem]{ulem}

\usepackage{graphicx}
\usepackage{amssymb} 
\usepackage{amsmath}
\usepackage{epstopdf}

\begin{document}

\begin{frontmatter}

\title{Photoconversion in HIT solar cells: Theory vs experiment}

 \author[label1]{A.V.~Sachenko}
 \author[label1]{Yu.V.~Kryuchenko}
 \author[label1]{V.P.~Kostylyov}
\author[label2]{A.V.~ Bobyl}
\author[label2,label3]{E.I.~Terukov}
\author[label3]{S.N.~Abolmasov}
\author[label3]{A.S.~Abramov}
\author[label3]{D.A.~Andronikov}
\author[label2]{M.Z.~Shvarts}
\author[label1]{ I.O.~Sokolovskyi}
\author[label4]{M.~Evstigneev\corref{cor1}}
 \address[label1]{V. Lashkaryov Institute of Semiconductor Physics, NAS of Ukraine, 41 prospect Nauky, 03028 Kyiv, Ukraine}
\address[label2]{Ioffe Institute, 194021 St.-Petersburg, Russian Federation}
\address[label3]{TFTC Ioffe R\&D Center, 194021 St.-Petersburg, Russian Federation}
\address[label4]{Department of Physics and Physical Oceanography, Memorial University of Newfoundland, St. John's, NL, A1B 3X7  Canada}
\cortext[cor1]{Corresponding author: mevstigneev@mun.ca}

\begin{abstract}
Theoretical expressions for the photocurrent in Heterojunction solar cells with Intrinsic Thin layer (HIT cells) are derived taking into account tunneling of electrons and holes through wide-bandgap layers of $\alpha$-Si:H or $\alpha$-SiC:H. The criteria, under which tunneling does not lead to the deterioration of solar cell characteristics, in particular, to the reduction of the short-circuit current and open-circuit voltage, are introduced. An algorithm to compute the photoconversion efficiency of HIT elements taking into account the peculiarities of the open-circuit voltage generation, in particular, its rather high values, is proposed. To test the theoretical predictions against the experimental results, HIT elements with the efficiency of about $20\,\%$ are manufactured, and their short-circuit current, open-circuit voltage, photoconversion power, and fill factor of the current-voltage curve are measured as a function of temperature in a wide temperature range from 80 to 420\,K. At low temperatures, the open-circuit voltage and the photoconversion power decrease as temperature is reduced. At $T \ge 200$\,K, the theoretical expressions and the experimental curves agree rather well. The behavior of the fill factor and output power at low temperatures is explained by the increase of the series resistance on cooling. The reasons behind the reduction of the power temperature coefficient in HIT elements are discussed and shown to be related to the low surface and volume recombination rates. Finally, a theoretical expression for the HIT element's temperature under natural working conditions is derived.
\end{abstract}

\begin{keyword}
High-efficiency solar cells \sep Crystalline silicon \sep Heterojunction
\PACS 88.40.hj \sep 88.40.fc \sep 73.40.Lq \sep 73.50.Pz \sep 73.40.Gk
\end{keyword}
\end{frontmatter}


\section{Introduction}
In the last years there has been substantial progress in increasing the photoconversion efficiency of the so-called HIT (Heterojunction with Intrinsic Thin layer) solar cells. For instance, the work \cite{Mas14} reports the efficiency of 25.6\,\% for AM1.5 spectrum. However, according to the estimates performed in several works, such as \cite{Sho61}, the highest theoretical efficiency of Silicon-based photoconverters under unfocused solar radiation is about 30\,\%. In order to approach the limit efficiency, it is necessary to intensify the research aiming at optimizing the parameters of these structures. This requires deeper understanding of the peculiarities of the processes that take place in HIT elements as compared to the standard solar cells.

The principal difference between the HIT elements and the p-n junction-based solar cells is that, in order to contribute to the photoinduced current, the photogenerated electrons and holes need to travel through the wide-bandgap layers of $\alpha$-Si:H or $\alpha$-SiC:H \cite{Mas14, Suw10}. The band diagram of a heterojunction depends both on the bandgaps of the semiconductors in contact and on the respective electron affinities. The absence of a barrier on the heterojunction boundary for at least one band is not a rule, but an exception. In the case considered, the very thin layers of wide-bandgap semiconductors behave as dielectrics. The current transport through such layers proceeds via either direct or multistep trap-assisted tunneling \cite{Vul83}. As estimates show, the former mechanism cannot produce the transfer of photogenerated electrons and holes for dielectric thicknesses of the order of 10\,nm, the typical thickness of the $\alpha$-Si:H or $\alpha$-SiC:H layers in HIT elements. Therefore, the photocurrent is most likely due to the trap-assisted tunneling.

In this work, we have manufactured HIT elements with the efficiency close to 20\,\% and measured their main characteristics, such as the short-circuit current, $I_{SC}$, the open-circuit voltage, $V_{OC}$, the fill factor, $FF$, and the photogenerated power, $P$, in a broad temperature range from 80 to 420\,K. 

In order to explain theoretically the results obtained, we derive the conditions, under which the determination of the photoconversion efficiency $\eta$ in the HIT elements can be performed using the same procedure as in the usual p-n junction-based solar cells. The model presented here takes into account the distinct features of the HIT solar cells as compared to the conventional solar cells. In the first place, these features are related to the high bulk Shokley-Hall-Reed lifetime $\tau_{SR}$ ($\ge 1$\,ms). Due to this, the inequalities $\Delta p \ge N_d$ and $L \gg d$ hold in a broad temperature range, where $\Delta p$ is the excess electron-hole density in the base region (Si single crystal), $N_d$ the doping level, $L$ the diffusion length of the minority carriers in the base, and $d$ the base thickness. We show that for $\Delta p \ge N_d$, the open-circuit voltage $V_{OC}$ is higher than in the standard case $\Delta p < N_d$ not only because of the higher $\Delta p$-value, but also because of the additional contribution due to the back surface of the solar cell.

We take into account several recombination mechanisms. We establish that in HIT solar cells at $T > 200$\,K, Shockley-Hall-Reed and surface recombination dominate  at relatively low base doping level, and interband Auger recombination at high base doping level. We explain the notable insensitivity of the HIT elements' photoconversion efficiency to temperature as compared to the commercial solar cells \cite{Mis11, Sko09}, first and foremost, by the essential reduction of the total bulk and surface recombination rates in the HIT elements.

In order to explain the experimental temperature dependence of the fill factor and photoconversion power, we take into account the increase of the series resistance in the low-temperature range, which is due to the increase of the contact resistance.

Finally, by means of solving the equations for photogenerated current and voltage coupled to the temperature balance equation, we find the temperature of HIT element under working conditions. We show that an element operating under the AM0 radiation conditions in the environment with the temperature of about 173\,K (-100\,$^\circ$C) has the temperature exceeding 300\,K. The reason is insufficient radiative cooling, which, in turn, is related to the fact that about 80\,\% of the solar energy incident on the HIT elements is converted into heat. Under the AM1.5 conditions, cooling down of the solar cells is due to radiation and convection, as a result of which the temperature of the solar cell notably decreases. However, apart from the cases of high wind velocity, the temperature of the solar cells under natural conditions exceeds the environment temperature by about 10\,$^\circ$C.

\begin{figure}[t!] 
\includegraphics[scale=0.3]{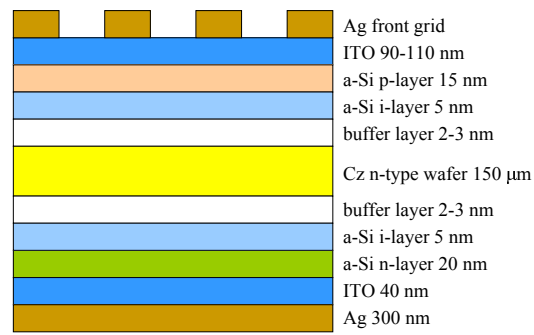}
\caption{Schematic illustration of our HIT-element.}
\label{fig1}
\end{figure}

\begin{table*}[t!]
\centering

\label{table1}
\caption{Parameters of our HIT element. $I_{SC}$, $V_{OC}$, and $P$ are given for $T = 300$\,K.}
\begin{tabular}{| l | l | l | l | l | l | l | l | l | l | l |}
\hline
$N_d$, cm$^{-3}$ &  $d$, $\mu$m & $\tau_b$, ms & $I_{SC}$, mA & $V_{OC}$, V & $S$, cm/s & $P$, W & $R_{S0}$, Ohm  & $A_t$, $\mu$Ohm & $\varphi_c$, V & $E_{00}$, meV\\ \hline
10$^{15}$ & 150 & 1 & 158.6 & 0.682 & 25 & 0.08 & 0.2 & 0.4 & 0.39 & 23\\
\hline
\end{tabular}
\end{table*}

\section{Fabrication and characterization of HIT elements}
The structure of our HIT element is shown in Fig.~\ref{fig1}. It contains a single-crystal  n-type c-Si substrate of (001) orientation, produced by Czochralski method (Cz). It has the doping level $N_d \approx 10^{15}$\,cm$^{-3}$, thickness $d \approx 150\,\mu$m, and Shockley-Hall-Reed lifetime of charge carriers $\tau_{SR} \approx 1$\,ms. On its front side we deposited a buffer layer followed by layers of intrinsic amorphous silicon $\alpha$-Si, amorphous p-type $\alpha$-Si, conducting layer of transparent indium tin oxide (ITO), and a silver grid that collects current. On the back side are a buffer layer, followed by layers of intrinsic $\alpha$-Si, n-type $\alpha$-Si, transparent conducting ITO, and silver. The buffer films, produced by plasma-enhanced chemical vapor deposition, ensure passivation of the surface states and effective reduction of recombination losses related to the surface recombination rate. The thicknesses of these layers are of the order of 10\,nm. The parameters of our HIT elements are indicated in Table~\ref{table1}.

The last fabrication step was deposition of a solid contact film on the back side and contact grid on the front side of the element. The photoactive surface area of our HIT elements, including the contact grid, was 4.34\,cm$^2$.

In order to facilitate absorption of the incident radiation by the semiconductor bulk on both sides of the substrate, we created microtextures in the form of randomly located vertical pyramids with characteristic size of a few micrometers. In the work \cite{Mas14} it was shown that such surface morphology allows one to obtain the short-circuit current density $J_{SC}$ of 41.8\,mA/cm$^2$ under AM1.5 conditions. Estimates that account for shadowing of the solar cell by the contact grid indicate that, in this case, the external quantum yield is close to 96\,\%, whereas the internal quantum yield of the photogenerated current is close to 100\,\%.

\section{Theoretical formulation}
The mathematical description of photoconversion processes in HIT elements is similar to the already solved problem of photoconversion efficiency in structures consisting of a conducting layer, a tunneling layer of SiO$_2$, and a layer of Si, where the majority carriers (electrons) enter the metal via thermionic emission, and the minority carriers (holes) tunnel through the dielectric layer \cite{Vul83}. In HIT elements, both electrons and holes  tunnel through the thin dielectric layer. 

Taking into account tunneling through the wide-bandgap layers and the fact that the density of surface states is extremely small, the boundary conditions for the electron and hole current densities, $J_n$ and $J_p$, in the case of non-degenerate semiconductor  read:
\begin{eqnarray}
&&J_p = q\frac{V_p}{4}\theta_p\left[p(x) - p_0(x)\right]_{x = 0}\ , \nonumber \\
&&J_n = q\frac{V_n}{4}\theta_n\left[n(x) - n_0(x)\right]_{x = d}\ .
\label{1}
\end{eqnarray}
Here, $q$ is the elementary charge, $V_p$ and $V_n$ are the mean thermal velocities of holes and electrons, respectively, $\theta_p$ and $\theta_n$ are their tunneling coefficients, $p(x)$ and $n(x)$ are the densities of electrons and holes, and $p_0(x)$ and $n_0(x)$ are the respective equilibrium densities. In the second expression, $d$ is the thickness of Si layer.

As shown in \cite{Vul83}, the minority carriers' current (for definiteness, we assume these to be holes) is collected without recombination losses if the condition
\begin{equation}
V_{pe} = \frac{V_p\theta_p}{4}e^{-y(0)} \gg V_r + S\ ,
\label{2}
\end{equation}
is fulfilled, where $V_{pe}$ is the effective hole emission rate from the semiconductor into the metal, $y(0)$ is the nonequilibrium electrostatic potential (band bending) normalized to $k_BT/q$ in the $x = 0$ plane, $V_r$ and $S$ are, respectively, bulk and surface recombination rates. The inequality (\ref{2}) should hold not only in the short-circuit regime, but also in the regime of maximal power. Estimates obtained for the typical parameter values of the HIT elements indicate that (\ref{2}) is fulfilled for $\theta_p \ge 10^{-5}$, which is the case in our HIT elements.

The dependence of the short-circuit current on temperature is due to the temperature dependence of the bandgap and absorption coefficent \cite{Spa78}. It can be expressed as
\begin{equation}
I_{SC}(T) = q \int_{\lambda_0}^{\lambda_m(T)}d\lambda\,Q_e(\lambda, T)\,\Phi(\lambda)\frac{\lambda}{hc}\ ,
\label{3a}
\end{equation}
where $\lambda_m(T) = hc/E_g(T) \approx 1.24\,\mu\text{m\,eV}/E_g(T)$ is the photoelectric threshold in Si, $\lambda_0$ is the absorption edge, $Q_e(\lambda, T)$ is the external quantum efficiency of the cell, and $\Phi(\lambda)$ is the spectral density of the incident radiation. The temperature dependence of the bandgap in Si can be found, e.g., in \cite{Gre90}. The increase of $I_{SC}(T)$ with temperature is due to the variations of both $\lambda_m(T)$ and $Q_e(\lambda, T)$ with $T$. Because the latter dependence is difficult to determine theoretically, we use, instead of Eq.~(\ref{3a}), an empirical expression, which approximates the product $Q_e\Phi$ in the integrand of (\ref{3a}) by a black-body spectrum:
\begin{equation}
I_{SC}(T) = I_{SC}(300\,\text{K})\,\frac{F(T)}{F(300\,{\text{K})}}\ ,
\label{9}
\end{equation}
where 
\begin{equation}
F(T) = \int_0^{\lambda_m(T)}\frac{d\lambda}{\lambda^4\left(e^{hc/(kT_{eff}\lambda)}-1\right)}\ .
\end{equation}
In general, the effective temperature $T_{eff}$ is smaller than the radiation temperature $T_L$. However, under the AM0 conditions, or when an incandescent lamp is used, the two temperatures are practically the same, provided that the quantum efficiency is close to 1.

For the open-circuit voltage in HIT elements, the following general expression applies:
\begin{eqnarray}
&&V_{OC} = \frac{k_BT}{q}\Big(\big(\Delta y_{OC}(0) - \Delta y_{SC}(0)\big)\nonumber\\
&&\ \ \ \ \ \ \ \ \ \ \ \  - \big(\Delta y_{OC}(d) - \Delta y_{SC}(d)\big)\Big)\ ,
\label{3}
\end{eqnarray}
where $\Delta y_{OC}(0)$ and $\Delta y_{SC}(0)$ are the dimensionless variations of the band bending in the open-circuit and short-circuit regimes, respectively, at $x = 0$, and $\Delta y_{OC}(d)$ and $\Delta y_{SC}(d)$ are the respective values at $x = d$.

Our estimates indicate that in HIT elements that completely conduct the  current generated in the short-circuit regime, the following conditions hold: $\Delta y_{SC}(0) \ll 1$ and $-\Delta y_{SC}(d) \ll 1$ (see Section~\ref{appendix}). Therefore, in this case, the value of $V_{OC}$ is practically the same as in the standard p-n junction-based solar cells. Taking into account that $p_0 n_0 = n_i^2$, where $n_i$ is the intrinsic charge carrier density, the open-circuit voltage is
\begin{eqnarray}
&&V_{OC} \approx V_{OC1} + V_{OC2}\ ,\ V_{OC1} = \frac{kT}{q}\ln\frac{\Delta p_0 n_0}{n_i^2} \nonumber\\
&&
V_{OC2} = \frac{kT}{q}\ln\left(1 + \frac{\Delta p_0}{n_0}\right)\ ,
\label{4}
\end{eqnarray}
where $\Delta p_0$ is the excess electron-hole pair density in the Si bulk in the open-circuit regime.
The expression (\ref{4}) for $V_{OC}$ is valid when the moduli of band bending at $x = 0$ and $x = d$ in the absence of illumination exceed the respective values in the open-circuit regime,
\begin{equation}
y_0(0) > qV_{OC1}/kT\ ,\ \ y_0(d) > qV_{OC2}/kT\ .
\label{5a}
\end{equation} 
Usually, these inequalities are well satisfied in HIT elements, at least at the operating temperatures.

Eq.~(\ref{4}) is a quadratic equation for $\Delta p_0$ with the solution
\begin{equation}
\Delta p_0 = -\frac{n_0}{2} + \sqrt{\frac{n_0^2}{4} + n_i^2e^{qV_{OC}/k_BT}}\ .
\label{5}
\end{equation}
The open-circuit voltage can be found from the balance equation relating the short-circuit current and the recombination currents due to the Shockley-Hall-Reed mechanism with the characteristic time $\tau_{SR}$, radiative recombination with the time $\tau_r$, Auger recombination with the rate $R_{Auger}$, surface recombination with the rate $S$, and recombination in the space-charge region with rate $R_{SC}$:
\begin{equation}
I_{SC} = qA_{SC}\left[d\left(\tau_{SR}^{-1} + \tau_r^{-1}\right) + R_{Auger} + S + R_{SC}\right]\Delta p_0
\label{6}
\end{equation}
where \cite{Han90, Sac07}
\begin{eqnarray}
&&R_{Auger} = C_p(n_0 + \Delta p_0)\Delta p_0 + C_n(n_0 + \Delta p_0)^2\ ,\nonumber\\
&&C_p = 10^{-31}\,\text{cm}^6/\text{s}\ ,\nonumber \\
&&C_n = \left(2.8\cdot 10^{-31} + \frac{2.5\cdot 10^{-22}}{(n_0 + \Delta p_0)^{0.5}}\right)\,\text{cm}^6/\text{s}\ .
\label{7}
\end{eqnarray}
Here, $A_{SC}$ is the surface area of the solar cell. The space-charge region recombination rate is given by \cite{Sac15}
\begin{eqnarray}
&&R_{SC}(\Delta p_0) \approx \frac{L_D}{\tau_{SR}}\int_{y_{pn}}^{-0.1}dy
\frac{n_0 + \Delta p_0}{\sqrt{e^y - y - 1}}\nonumber\\
&&\ \ \ \times\Big[(n_0 + \Delta p_0)e^y + n_i(T)e^{\varepsilon_r} + \nonumber \\
&&\ \ \ b\left(\frac{n_i(T)^2}{n_0 + \Delta p_0} + \Delta p_0\right)e^{-y} + n_i(T)e^{-\varepsilon_r}\Big]^{-1}\ ,
\end{eqnarray}
where $L_D = \sqrt{\varepsilon_0\varepsilon_S kT/(2q^2n_0)}$ is Debye screening length, $\varepsilon_0$ the relative dielectric constant of the semiconductor, $b = \sigma_p/\sigma_n$ the ratio of the capture cross-sections of a hole and an electron by a deep recombination center, $\varepsilon_r = E_r/kT$ the normalized energy of the deep recombination level measured from the middle of the band gap, $y$ the dimensionless potential, and $y_{pn}$ the dimensionless potential at the p-n junction boundary.

The intrinsic charge carrier density in Si is given by the empirical expression \cite{Gre90}
\begin{equation}
n_i(T) = 5.71\cdot 10^{19} \left(\frac{T}{300\,{\text K}}\right)^{2.365}\,e^{-6773\,{\text K}/T}\,\text{cm}^{-3}\ ,
\label{a}
\end{equation}
and the equilibrium electron density can be obtained from the relations $n_0 = p_0 + N_d$, $p_0 n_0 = n_i^2$:
\begin{equation}
n_0 = \frac{N_d}{2} + \sqrt{\frac{N_d^2}{4} + n_i^2}\ ,
\end{equation}
where $N_d$ is the donor density.

As our estimates using the parameters from Table~\ref{table1} has shown, in the temperature range from 200\,K to 300\,K, $\Delta p_0$ is about $5\cdot 10^{15}$\,cm$^{-3}$, and $R_{SC}$ does not exceed 0.1\,cm/s. Thus, the space-charge region recombination rate is insignificant in comparison to the total bulk and surface recombination.

The fill factor, $FF$, can be written as
\begin{equation}
FF = \frac{I_mV_m}{I_{SC}V_{SC}}\left(1 - \frac{R_SI_m}{V_m}\right)\ ,
\label{10}
\end{equation}
where $I_m$ and $V_m$ are the photoinduced current and voltage in the maximum-power regime, and $R_S$ is the series resistance of the solar cell.

To determine $I_m$ and $V_m$ of a HIT element, one has to consider the current-voltage relation [cf. Eq.~(\ref{6})],
\begin{equation}
I(V) = I_{SC} - qA\big(d(\tau_{SR}^{-1} + \tau_r^{-1}) + R_{Auger} + S\big)\,\Delta p(V)\ , 
\label{11}
\end{equation}
where 
\begin{equation}
\Delta p(V) = -\frac{n_0}{2} + \sqrt{\frac{n_0^2}{4} + n_i^2e^{qV/k_BT}}\ .
\label{12}
\end{equation}
The maximum power condition $d(VI(V))/dV = 0$ yields the value of $V_m$, whose substitution into (\ref{11}) allows one to determine $I_m$.

The total series resistance $R_S$ consists of the bulk resistance of the material, $R_{bS}$, sheet resistance $R_{SS}$ of the ITO film used as a transparent conductor, and contact resistance $R_{CS}$. The first two contributions to series resistance are usually temperature-independent. The temperature dependence of the contact resistance is determined by the nature of the current transport mechanism through the contact. If the current is due to thermionic or field emission, $R_{CS}$ increases on cooling. The resistance $R_{CS}$ is presumably related to the resistance of ITO-metal contact. Because the ITO film is degenerate, we can expect that the current transport through the contact is due to the field emission mechanism. Then, the expression for the series resistance can be written as \cite{Sze}
\begin{equation}
R_S = R_{S0} + A_t(T)\exp\left(\frac{q\phi_C}{E_{00}\coth(E_{00}/k_BT)}\right)\ ,
\label{13}
\end{equation}
where $R_{S0}$ is the temperature-independent component of the series resistance, $A_t(T)$ the prefactor that depends on temperature according to a power law, $\phi_C$ is the contact barrier height, $E_{00}$ is the characteristic electron tunneling energy in the semiconductor. If the barrier $\phi_C$ is sufficiently high, then one can neglect the temperature dependence of the prefactor.

The photogenerated power can be found from the standard expression
\begin{equation}
P = I_{SC}V_{OC}FF
\label{14}
\end{equation}
and the photoconversion efficiency is given by
\begin{equation}
\eta = \frac{P}{A_{SC}P_S}\ ,
\label{15}
\end{equation}
where $P_S$ is the surface power density of the incident radiation.

\begin{figure}[t!] 
\includegraphics[scale=0.3]{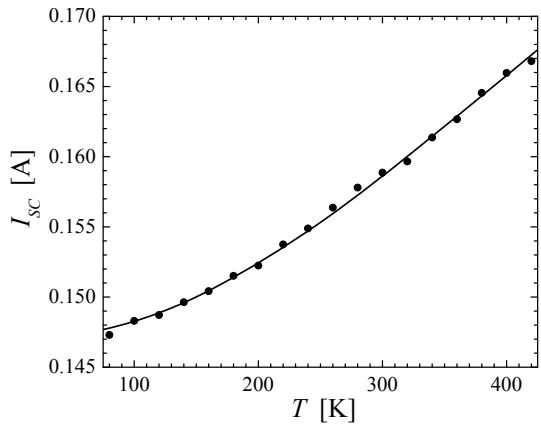}
\caption{Experimental (circles) and theoretical (line) temperature dependence of the short-circuit current.}
\label{fig2}
\end{figure}

\section{Comparison between the theoretical and experimental characteristics of HIT elements}
\label{sec4}
Fig.~\ref{fig2} shows the short-circuit current vs. temperature curve obtained using a xenon lamp as illumination source. The theoretical curve was produced by fitting the experimental data points with Eq.~(\ref{9}); the best fit was achieved for the effective temperature $T_{eff} = 4030$\,K. The agreement of the experimental and the theoretical curve in the whole temperature range is obvious.

\begin{figure}[t!] 
\includegraphics[scale=0.3]{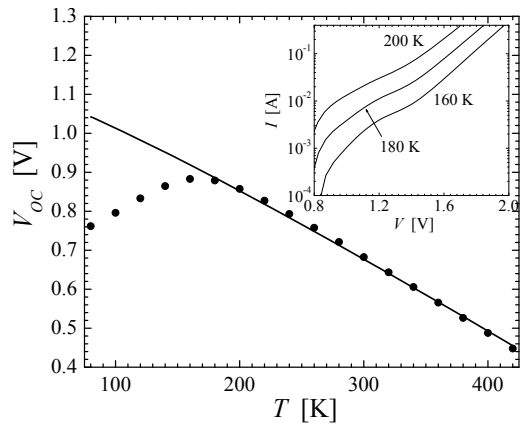}
\caption{Experimental (circles) and theoretical (lines) temperature dependence of the open-circuit voltage  obtained using the general expression (\ref{6}) with parameters shown in Table~\ref{table1}. The inset shows the current-voltage curves at three different temperatures in the low-temperature range.}
\label{fig3}
\end{figure}

Shown in Fig.~\ref{fig3} is the temperature dependence of the open-circuit voltage, $V_{OC}(T)$. The theoretical curves are obtained by numerically solving Eq.~(\ref{6}) supplemented by Eqs.~(\ref{7}) and (\ref{9}) with parameters from Table~\ref{table1}. When building these curves, we used the temperature-dependent $I_{SC}(T)$, and the intrinsic carrier density $n_i(T)$ given by Eq.~(\ref{a}). We neglected the temperature dependence of the Shockley-Hall-Reed, radiative, surface, and Auger recombination rates, because they are much weaker than the $n_i(T)$ dependence. It is noteworthy that for the equilibrium electron densities $n_0 \simeq 10^{15}$\,cm$^{-3}$, which correspond to the doping level, interband Auger recombination and radiative recombination can be neglected in comparison to the bulk Shockley-Hall-Reed and surface recombination. Auger recombination becomes essential when $n_0 \ge 10^{16}$\,cm$^{-3}$.

As can be seen in Fig.~\ref{fig3}, the theoretical curve agrees well with the experimental results at $T \ge 180$\,K. The temperature coefficient of $V_{OC}(T)$ at $T > 200$\,K is about 1.8\,mV/K, which is close to the typical value of $|dV_{OC}/dT| \approx 2$\,mV/K in the industrial silicon-based solar cells with graded p-n junctions. At temperatures below 180\,K, the experimental and theoretical values of $V_{OC}(T)$ diverge: the theoretical open-circuit voltage increases, whereas the experimental one decreases on cooling. 
 
Finding possible reasons for this non-monotonicity is a complicated problem. First of all, as follows from the calculations based on the spectrum of localized donor-like and acceptor-like states in $\alpha$-Si:H, this material is a heavily compensated semiconductor \cite{Kry15}, whose degree of compensation increases on reducing the temperature. Therefore, the charge in the acceptor-doped $\alpha$-Si:H decreases on cooling, leading to the reduction of $y(x = 0)$. This implies that, although the criterion (\ref{2}) at $T \le 200$\,K is fulfilled in the short-circuit regime, it breaks down at a relatively weaker bias. This point is illustrated by the experimental current-voltage curves of our HIT elements measured at $T = 200, 180$, and 160\,K, see insert in Fig.~\ref{fig3}. Similar to \cite{Don66}, these curves exhibit sections due to the hole tunneling from the narrow-band single-crystal Si into $\alpha$-Si:H for relatively high direct bias ($\ge 1.4$\,V), whereas for weaker bias ($\le 0.82$\,V), the dark current-voltage curves are adequately described by the expressions of the form (\ref{11}) and (\ref{12}) (with $I_{SC} = 0$). Thus, the reduction of $V_{OC}(T)$ on cooling is due to several reasons. It is the decrease of $y(x = 0)$ and failure of the criterion (\ref{2}), i.e. the reduction of the photogenerated current at low temperatures. As the value of $y(x=0)$ goes down on cooling, the inequality $kTy_0(x = 0) > qV_{OC}$ breaks down, where $y_0(x = 0)$ is the dimensionless band bending in the absence of illumination. Then, the expression (\ref{4}) becomes inadequate, and the open-circuit voltage becomes smaller than the value predicted by (\ref{4}). Finding a better formula for $V_{OC}$ in the low-temperature range is a complicated task, which requires the knowledge of the distribution of the localized states in the doped and undoped layers of $\alpha$-Si:H. Additional research is needed to attack this problem.

\begin{figure}[t!] 
\includegraphics[scale=0.3]{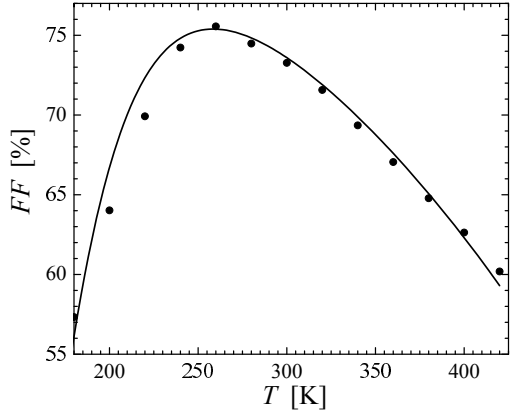}
\caption{Experimental (circles) and theoretical (lines) temperature dependence of the current-voltage curve fill factor.}
\label{fig4}
\end{figure}

Fig.~\ref{fig4} shows the temperature dependence of the fill factor. It is seen that $FF$ at low temperatures becomes anomalously low, signaling a strong increase of the series resistance $R_S$ on cooling. The theoretical curve was obtained from Eqs.~(\ref{10}) and (\ref{13}) for the same parameters as the theoretical curve $V_{OC}(T)$ from Fig.~\ref{fig3}. We took $A_t = 4\cdot 10^{-7}$ Ohm, $\varphi_c = 0.387$\,V, $E_{00} = 0.0225$\,eV, $R_{S0} = 0.2$\,Ohm. The agreement between the theoretical and the experiment curves is very good at $T \ge 200$\,K. 

\begin{figure}[t!] 
\includegraphics[scale=0.3]{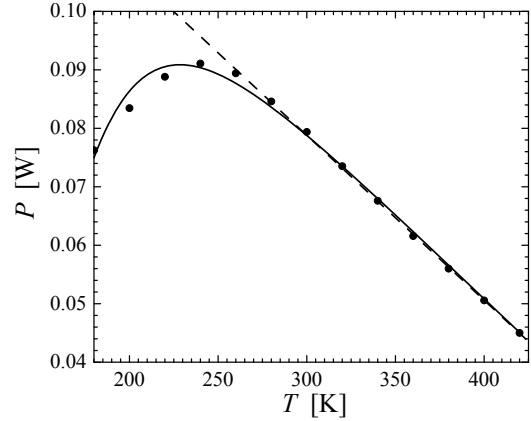}
\caption{Experimental (circles) and theoretical (lines) temperature dependence of the photoconversion power.}
\label{fig5}
\end{figure}

\begin{figure}[t!] 
\includegraphics[scale=0.3]{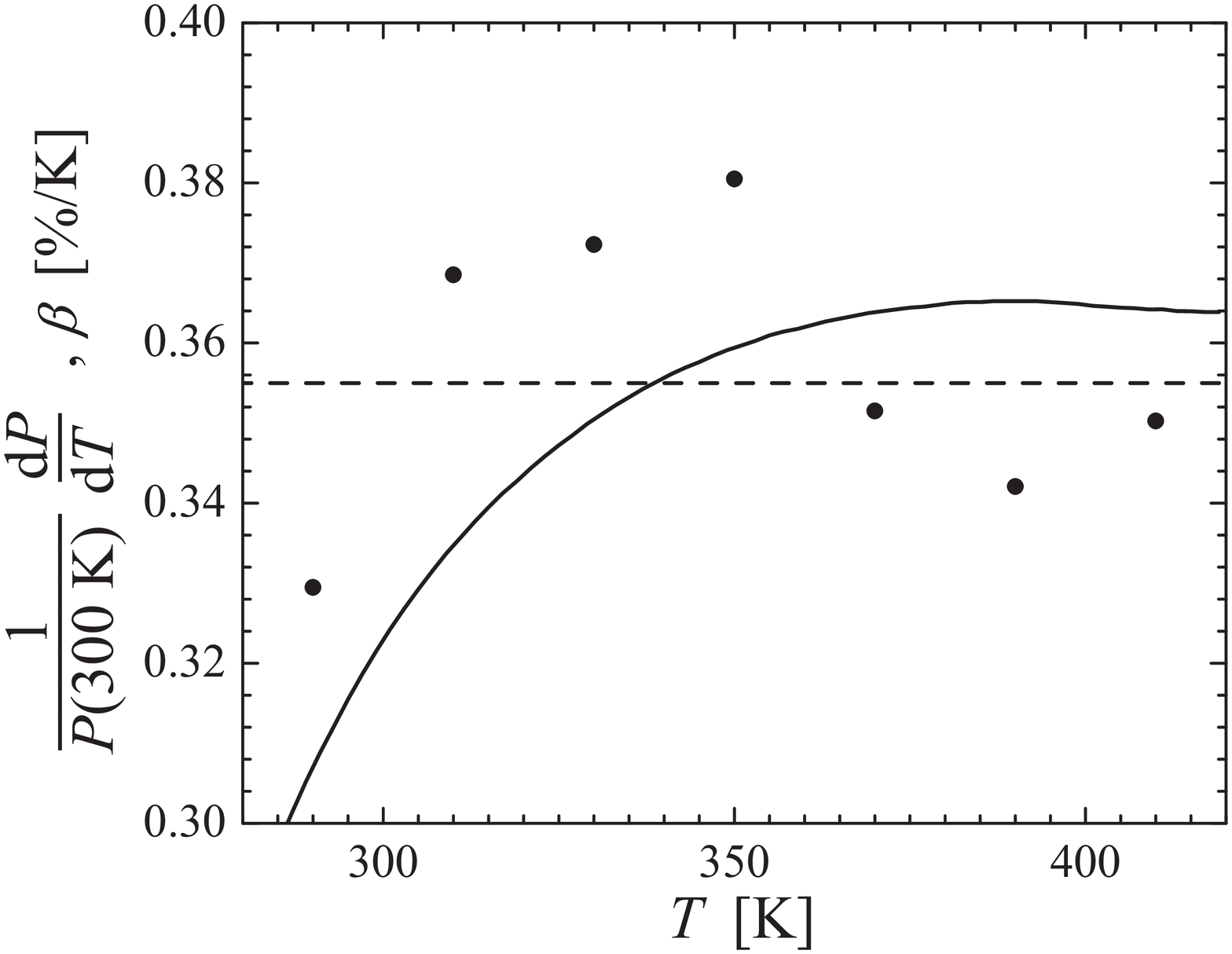}
\caption{Experimental (dashed line, circles) and theoretical (solid line) temperature dependence of the photoconversion power reduction coefficient. The dashed line was obtained by fitting the $P(T)$ curve, Fig.~\ref{fig5}, at $T > 280$\,K. The circles are obtained by numerically differentiating the experimental data using Eq.~(\ref{220}); the solid line is obtained from the derivative of the respective theoretical curve.}
\label{fig6}
\end{figure}

The temperature dependence of the photoconversion power $P$ in the maximal power regime, see Eq.~(\ref{14}), is shown in Fig.~\ref{fig5}. The agreement between theory and experiment at $T \ge 200$\,K, including its non-monotonic behavior, is obvious. Above about 280\,K, the maximal photoconversion power decreases approximately linearly with temperature. This linear decay can be described by an expression \cite{Sko09}
\begin{equation}
P(T) = P(300\,K)\left(1 - \beta (T - 300\,\text{K})\right)\ ,
\label{210}
\end{equation}
which defines the temperature coefficient $\beta$. A linear fit to the experimental data at $T \ge 280$\,K yields $\beta_{exp} = 0.355$\,\%/K, see Fig.~\ref{fig5}, dashed line. Theoretically, the photoconversion power vs. temperature curve slightly deviates from the linearity (\ref{210}), see the solid line in Fig.~\ref{fig6} showing the theoretical derivative of power with respect to temperature normalized to $P(300\,\text{K})$, as well as the respective experimental estimates. To obtain the experimental curve based on the set of experimental support points $(T_n, P_n)$, $n = 1, 2, \ldots$, we used the third-order accurate approximation for the derivative at the midpoint $T_{n+1/2} = (T_n + T_{n+1})/2$:
\begin{equation}
\frac{1}{P(300\,\text{K})}\frac{dP}{dT}(T_{n+1/2}) \approx \frac{1}{P(300\,\text{K})}\frac{P_{n+1} - P_n}{T_{n+1} - T_n}\ .
\label{220}
\end{equation}
The overall trent towards saturation of both experimental and theoretical curves is clear, and the discreapancy between the two sets of data is due to the amplification of the experimental uncertainty by the derivative expression (\ref{220}).

In the work \cite{Mis11}, the temperature coefficient of power reduction in HIT elements with $\eta = 23\,\%$ was measured. Its value at the largest open-circuit voltage $V_{OC}$ of 0.745\,V, which was realized at the smallest solar element thickness (see Fig.~3 of \cite{Mis11}), was 0.23\,\%/K. Our theoretical estimates at $T = 298$\,K and parameter values from \cite{Mis11}, in the thickness range of the Si substrate from 80 to 165\, $\mu$m, gives $\beta \approx 0.25\,\%$/K, quite close to the experimental values.

We note that the low values of $\beta(T)$ are directly related to the high excitation level in the open-circuit regime. This owes, in particular, to the minimal surface recombination rate in HIT elements. An important role here is played by atomic hydrogen, which leads to a strong reduction of the number of surface centers responsible for recombination. As shown in \cite{Yab86}, it was precisely the use of atomic hydrogen that allowed to obtain minimal surface recombination rate of about 0.25\,cm/s on Si.

The bulk recombination rate, $V_r = d/\tau_b$, where $\tau_b$ is the bulk lifetime, is also much smaller in HIT elements than in the graded p-n junction-based solar cells. This is so, because, as shown above, HIT elements employ silicon with long Shockley-Hall-Reed lifetimes exceeding 1\,ms. Secondly, $d$ in HIT elements is also smaller than in the usual solar cells. For instance, for $d = 100\,\mu$m and $\tau_{SR} = 1$\,ms, $V_r = 10$\,cm/s. These features make it possible to maintain a high excitation level $\Delta p_0 \gg n_0$ in HIT elements, and thus to maximally reduce the temperature dependence of $V_{OC}$ and $P$, and to minimize $\beta$.

\section{Operating temperature of the HIT elements}
In order to determine the element's temperature in a realistic case, one needs to amend the equations for the photogenerated current and voltage by the energy balance equation. For the radiative cooling mechanism, this equation was obtained in \cite{Sac14}. Taking the radiation and convection cooling mechanisms into account, it assumes the form:
\begin{equation}
P_S(1 - \varepsilon - \eta(T)) = aK_T\sigma(T^4 - T_S^4) + \gamma (T - T_S)\ .
\label{18}
\end{equation}
Here, $P_S$ is the power incident from the Sun, $\eta(T)$ the conversion efficiency of the cell, and the parameter $\varepsilon$ characterizes the energy dissipation due to radiative recombination. For silicon-based solar cells, where the Shockley-Hall-Reed lifetime, $\tau_{SR}$, as a rule, is much smaller than radiative recombination time, $\tau_r$, the value of the parameter $\varepsilon$ is close to 0. The value of the parameter $a$, which depends on the grayness of the element, i.e. on the closeness of the solar cell radiation spectrum to the black body spectrum, is also of the order of 1. The value of $K_T$ equals the ratio of the element's area that emits radiation to the area that receives radiation. The parameter $\sigma$ is the Stefan-Boltzmann constant, $T = T_S + \Delta T$ is the temperature of the solar cell, and $T_S$ is the environment temperature. Finally, $\gamma$ is the convection coefficient, which depends on the magnitude and direction of the wind velocity, air humidity, and barometric pressure \cite{Ros86}. 

\begin{figure}[t!] 
\includegraphics[scale=0.3]{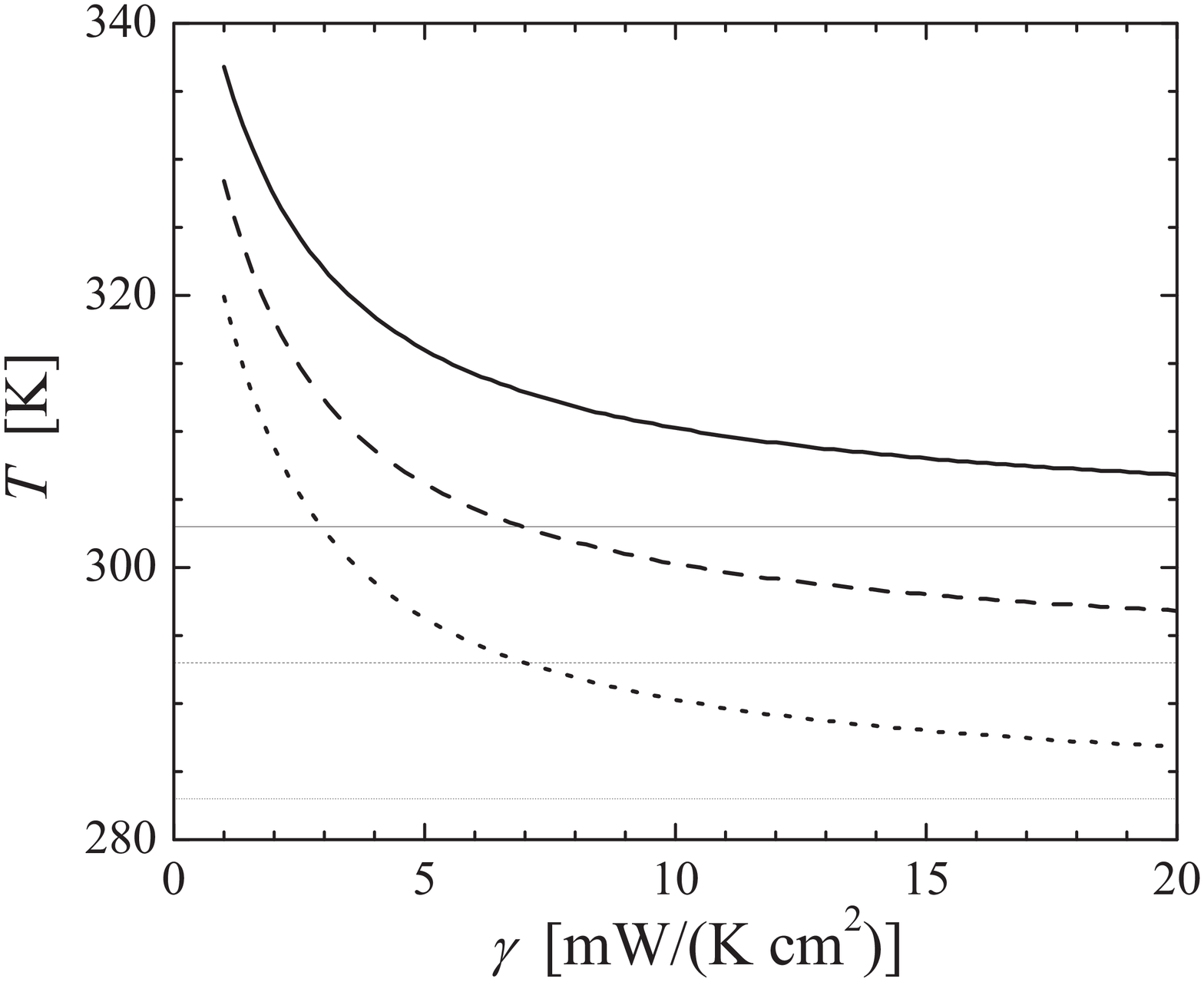}
\caption{Temperature of the HIT photovoltaic element vs. convection coefficient, $\gamma$. It is assumed that the surrounding temperature $T_S$ is 30, 20, and 10\,$^\circ$C (from top to bottom curves). The horizontal lines correspond to the environment temperatures.}
\label{fig7}
\end{figure}

Fig.~\ref{fig7} shows the results of the solution of Eqs.~(\ref{15}) and (\ref{18}) under the conditions of AM1.5 spectrum.

At high geographic latitudes, the main contribution to the yearly power yield of a solar plant comes from the time interval between spring and autumn, when the environment temperature is greater than 0\,$^\circ$C. The temperature of the solar cell is also positive. In this case, for the environment temperatures of 10, 20, and 30\,$^\circ$C and $\gamma = 6\cdot 10^{-3}$\,W/cm$^2$K, the theoretical operating temperature of the element exceeds the environment temperature by 11\,$^\circ$C, see Fig.~\ref{fig7}.

For the AM0 conditions, typically realized when the solar elements are placed on a satellite, the environment temperature is about 173\,K, see \cite{Dia06}. Solution of Eqs.~(\ref{15}) and (\ref{18}) with $\gamma = 0$ indicates that, in this case, because of the absence of convection mechanisms in the outer space, the working temperature of the solar cells is close to 49\,$^\circ$C.

\section{Conclusions}
Peculiarities of HIT as opposed to the standard solar cells include, on the one hand, tunneling of electrons and holes through the $\alpha$-Si:H or $\alpha$-SiC:H layers; on the other hand, large number densities of the excess electron-hole pairs and large diffusion length of the minority charge carriers. In this work, we have derived the criteria, under which tunneling processes do not lead to the deterioration of the HIT elements' characteristics as compared to those of the solar cells based on the graded p-n junctions, and obtained the expressions for the main parameters that characterize the performance of HIT elements. 

These expressions are compared to the experimental results. We have obtained good agreement between the theoretical and experimental temperature dependences of the HIT parameters in the temperature range $T \ge 200$\,K. At low temperatures, the series resistance, $R_S$, grows dramatically on cooling due to the contribution of the contact resistance. We would like to point out that the reduction of the open-circuit voltage, $V_{OC}$, fill factor, $FF$, and photoconversion power, $P$, at low temperatures is reported here for the first time and is interesting from the scientific point of view. The reduction of $FF$ and $P$ is explained by the reduction of $V_{OC}$, and by the increase of $R_S$. Of main practical interest are the results obtained for $T > 0\,^\circ$C. They are completely explained by the specifics of the HIT elements studies, especially the high excitation levels under AM1.5 conditions.

We have shown that the rather low value of the calculated temperature coefficient of photoconversion power reduction, $\beta(T) \approx 0.3$\,\%/K, near room temperature is in agreement with the data from \cite{Mis11}. The physical reason of its low values in HIT solar cells is low surface and bulk recombination rate as compared to the silicon graded p-n junction-based solar elements.

\begin{figure}[t!] 
\includegraphics[scale=0.3]{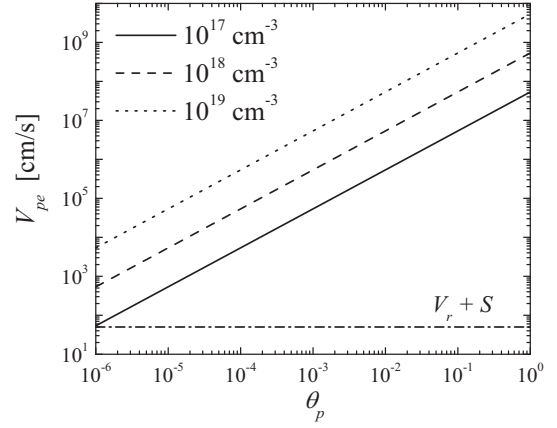}
\caption{Theoretical dependence of $V_{pe}$ on $\theta_p$ in the highest-power regime.}
\label{fig8}
\end{figure}

\begin{figure}[t!] 
\includegraphics[scale=0.3]{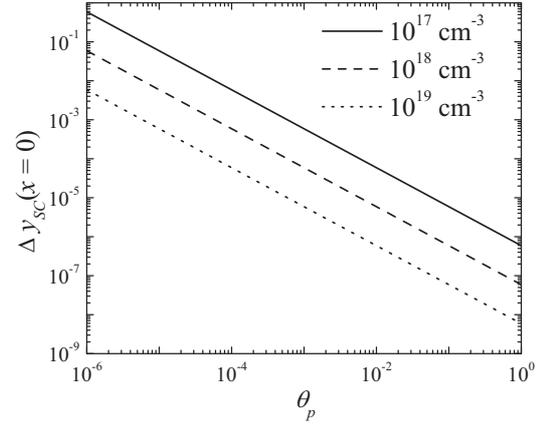}
\caption{Theoretical dependence of $\Delta y_{SC}(x = 0)$ on $\theta_p$.}
\label{fig9}
\end{figure}

\appendix
\section{Justification of the conditions $\Delta y_{SC}(0) \ll 1$, $|\Delta y_{SC}(d)| \ll 1$}
\label{appendix}
Shown in Fig.~\ref{fig8} is the dependence of $V_{pe}$ on $\theta_p$ in the maximal power regime at $T = 300$\,K. When building these plots, we took into account that the dimensionless band bending in the $x = 0$ plane is lower than the equilibrium value, i.e. $y(0) = y_0(0) - qV_m/k_BT$, where $V_m = 0.65$\,V is the photogenerated voltage in the maximal-power regime.  We have taken three values of the parameter $p_0(0) = p_0\,e^{-y_0(0)}$ of $10^{17}$, $10^{18}$, and $10^{19}$\,cm$^{-3}$. For all these values, conductivity inversion in Si is realized near the boundary at $x = 0$, i.e. a charge-induced p-n junction is formed in the p-Si:H layer, which is necessary for normal functioning of the solar cell. The value of $p_0 = 10^5$\,cm$^{-3}$ was taken, which corresponds to the doping level of $n_0 = 10^{15}$\,cm$^{-3}$. The total value of $S + V_r$ was taken to be 50\,cm/s. As seen in Fig.~\ref{fig8}, the minimal value of $V_{pe}$ equal to $S + V_r$ is realized at $\theta_p = 10^{-6}$ for $p_0(0) = 10^{17}$\,cm$^{-3}$. For $\theta_p \ge 10^{-5}$, $V_{pe}$ notably exceeds $S + V_r$ for all three values of $p_0(0)$ tested, i.e. the criterion (\ref{2}) is well fulfilled.

In the Si case, the quasi Fermi levels of holes and electrons in the space-charge regions near the surface are, as a rule, constant. For non-degenerate semiconductors, the relations hold:
\begin{equation}
p = (p_0 + \Delta p)e^{-y(0)}\ ,\ \ n = (n_0 + \Delta p)e^{y(d)}\ ,
\label{1A}
\end{equation}
where $p_0$ and $n_0$ are the equilibrium electron and hole densities, and $\Delta p$ is the excess electron-hole pair density in the Si neutral region.

Here and in the following we assume that the electron-hole pair diffusion length in HIT elements, $L = \sqrt{D_p\tau_b}$ ($D_p$ is hole diffusion coefficient, $\tau_b$ the lifetime), is much greater than the thickness of Si wafer, i.e. $L \gg d$. Because of this, $\Delta p$ is the same at all $x$.

In the short-circuit regime
\begin{equation}
J_p = J_n = J_{SC}\ ,
\label{3A}
\end{equation}
where $J_{SC}$ is the short-circuit current density. Substitution of the expressions (\ref{1}) into this identity shows that the reduction of the coefficients $\tau_p$ and $\tau_n$ leads to the increase of the difference between the electron and hole densities on the contacts, $p(0) - p(d)$ and $n(0) - n(d)$, inversely proportionally to to $\theta_n$ and $\theta_p$. The estimates for $J_{SC} = 40$\,mA/cm$^2$ indicate that for $\theta_{p,n} = 10^{-5}$, these differences are about $10^{16}$\,cm$^{-3}$. Although this is a rather high value, it is substantially smaller than the equilibrium densities $n_0$ and $p_0$. This is called the accumulation effect, first pointed out at in \cite{Gut76}. Due to the accumulation effect, upon turning the illumination on, the short-circuit current saturates over the time scale necessary for the build-up of the difference $p(0) - p_0(0))$, at which the tunneling current becomes equal to the generation current.

We note that the relation (\ref{3A}) and the expression
\begin{equation}
\Delta p_{SC} = \frac{J_{SC}}{q V_{pe}}\ ,
\end{equation}
which is valid under the criterion (\ref{2}) in the short-circuit regime allow one to find the dimensionless bend bending in the $x = 0$ plane, $\Delta y_{SC}(0)$. In particular, if $\Delta y_{SC}(0) \ll 1$, we have
\begin{equation}
\Delta y_{SC}(0) = \frac{J_{SC}}{qp_0(0)(V_p\theta_p/4)}\ .
\end{equation}

Fig.~\ref{fig9} shows $\Delta y_{SC}(0)$ as a function of $\theta_p$ ad different values of $p_0(0)$. As seen in this figure, for $\theta_p > 10^{-6}$, the inequality $\Delta y_{SC}(0) \ll 1$ for all curves. That is, the criterion (\ref{2}), which ensures complete passage of the generation current through the wide-bandgap layer of $\alpha$-Si:H, also implies the smallness of $y_{SC}(0)$ in the short-circuit regime. We also note that if the coefficient $\theta_p$ and $\theta_n$ and the densities $n_0(d)$ and $p_0(0)$ are comparable, then, as calculations show, the criterion $\Delta y_{SC}(0) \ll 1$ also implies that $-\Delta y_{SC}(d) \ll 1$.

In the open-circuit regime, the relations (\ref{1}) and (\ref{1A}) allow to find the respective dimensionless potentials at $x = 0$ and $x = d$ as 
\begin{eqnarray}
&&\Delta y_{OC}(0) = y(0) - y_0(0) = \ln\frac{p_0 + \Delta p_0}{p_0}\ ,\nonumber \\
&&\Delta y_{OC}(d) = y(d) - y_0(d) = -\ln\frac{n_0 + \Delta p_0}{n_0}\ ,
\end{eqnarray}
where $\Delta p_0$ is the excess electron-hole pair density in the open-circuit regime, with $\Delta p_0 \gg n_0, p_0$. Substitution of these expressions into Eq.~(\ref{3}) yields the result (\ref{4}).

\section*{Acknowledgments}
This work was performed under the project of Minobrnauka RF  No. 14.607.21.0075 (RFMEFI60714X0075). M.E. is grateful to the Natural Sciences and Engineering Research Council of Canada (NSERC) for financial support.

\end{document}